\documentclass{article}
\usepackage{spconf,amsmath,graphicx,hyperref}

\usepackage{balance}
\usepackage{caption}
\usepackage{subcaption}
\usepackage{amsmath,graphicx,amssymb}
\usepackage[table]{xcolor}
\usepackage{multirow}
\usepackage[linesnumbered,ruled,vlined]{algorithm2e}
\usepackage{float} 
\SetKwInput{KwInput}{Input}                
\SetKwInput{KwOutput}{Output}              
\SetKwInput{KwModelParameters}{Model Parameters}   
\SetKwInOut{Kwrepeat}{repeat}

\title{Teaching the Teachers: Boosting unsupervised domain adaptation in speech recognition by ensemble update}
\name{Rehan Ahmad$^{1}$, Muhammad Umar Farooq$^{2}$\thanks{$^{2}$This work was done while at the university of Sheffield, UK}, Qihang Feng$^{1}$, Thomas Hain$^{1}$}

\address{%
  $^{1}$University of Sheffield, UK
  $^{2}$Emotech Ltd.
}

\begin{document}
\ninept
\maketitle
\begin{abstract}
Speech recognition systems often struggle with data domains that have not been included in the training. To address this, unsupervised domain adaptation has been explored with ensemble and multi-stage teacher-student training methods reducing the word error rate. Despite improvements, the error rate remains much higher than that achieved with supervised in-domain training. This work proposes a more efficient strategy by simultaneously updating the ensemble of teacher models along with the single student model eliminating the need for sequential models training. The joint update improves the word error rate of the student model, benefiting the progressively enhanced teacher models. Experiments are conducted with three labelled source datasets, namely AMI, WSJ, LS360, and one unlabeled target domain i.e. SwitchBoard. The results show that the proposed method improves the WER by 4.6\% on the Switchboard eval00 test set, thus outperforming multi-stage and iterative training methods.
\end{abstract}

\section{Introduction}
\label{sec:intro}

Automatic speech recognition (ASR) performance has improved significantly with advanced deep learning based models \cite{graves2012connectionist, li2022recent, baevski2020wav2vec}. However, previous studies 
\cite{likhomanenko21_interspeech, nguyen2020toward,zhu2023boosting, hsu21_interspeech} show that these models perform poorly when evaluated on out-of-domain (OOD) data. This mismatch between training and test domains is commonly found in real world situations  and can lead to significant performance degradation. Hence, domain adaptation methods \cite{bell2020adaptation} are of a great interest in ASR. Domain adaptation can be applied through supervised \cite{sim2018domain} or unsupervised \cite{khurana2021unsupervised} methods, depends on the availability of labels for the target domain. With labelled data, adaptation can be implemented by fine-tuning \cite{joshi2022simple} an already trained model to the target domain. However, it is more common to have no labels for the target domain. Unsupervised domain adaptation (UDA) is more challenging and requires carefully designed methods \cite{manohar2018teacher, khurana2021unsupervised} to adapt to a new domain. 

Teacher-student training (T/S) \cite{KD} is a well-known approach that has been applied to many speech applications, including domain adaptation \cite{asami2017domain, meng2019domain, zhang2020semi}. Many T/S based adaptation methods apply adaptation with parallel source and target data \cite{li2017large}, or with limited amount of labelled target data \cite{zhang2020semi} hence achieving better adaptation. On the other hand, unsupervised domain adaptation is challenging \cite{manohar2018teacher}, therefore their performance still lags behind creating a gap to improve such methods. Using an ensemble of teacher models in T/S training \cite{fukuda2017efficient, gao2021distilling, li2017semi} has shown great benefit to acoustic model training, as multiple teachers are found to provide complementary information. For example, work presented in \cite{DGKD} has shown the advantage of using an ensemble of teacher models instead of a single teacher for unsupervised domain adaptation. The method aims to select the best outputs from the teachers on unlabelled data to train a student model. The results show significant WER improvement of about 8.4\% absolute on an unlabelled conversational telephone speech using labelled data from read and meeting speech. This approach  was further improved in \cite{ahmad2024progressive} by sequential training of multiple student models \cite{rathod2022multi}. 

Many studies of iterative training for ASR \cite{khurana2021unsupervised, manohar2021kaizen, xu20b_interspeech, momentum} have shown that models can be improved by iteratively updating on the target domain by generating new labels in each iteration. In \cite{xu20b_interspeech}, authors have presented an iterative pseudo-labelling (IPL) technique with a single model, and showed its usefulness in a low resource setting where both source and target data are from the same domain. In \cite{momentum} and \cite{manohar2021kaizen} it was shown that a T/S training method outperforms iterative-pseudo labelling when both teacher and student models are iteratively updated on labelled and unlabelled data. The method presented in \cite{momentum} was tested on both in-domain and out-of-domain settings outperforming in both scenarios. The datasets were taken from read speech and conference talks having well articulated speech compared to conversational speech data, making model easier to adapt. 
Similarly, in \cite{manohar2021kaizen} authors experimented with only in-domain data for low-resource supervised data, outperforming other IPL methods.

This paper proposes to improve unsupervised domain adaptation by integrating the iterative method \cite{momentum, manohar2021kaizen} in ensemble T/S training \cite{DGKD}, where all the teacher models are updated simultaneously with the student model during training. Previous work on multi-stage training \cite{ahmad2024progressive} only considered previously trained student models as a teacher for next stage training. This work takes a significant and novel step in aiming not to loose the ensemble in sequential training of the students. The approach leads to improved adaptation because the teacher models progressively generate better labels for the student. In detail, teacher models are initially trained on source domain labelled data, where student model is initialized randomly. In each training iteration, the student model is updated with the unlabelled data from target domain, with labels acquired from the teachers. The teacher models are updated with an exponential moving average of the student model weights, making teachers update a simple inexpensive method. Experiments are conducted with three labelled data sources, consisting of meeting and read speech i.e. AMI \cite{Carletta2006}, WSJ \cite{paul1992design} and LS360 \cite{Panayotov2015}. The target unlabelled data is SwitchBoard \cite{Godfrey1992} which belongs to the telephone conversational domain. The results show that the proposed method outperform multi-stage \cite{DGKD} and iterative \cite{manohar2021kaizen} methods by significant WER improvement of about 4.6\%.

\let\thefootnote\relax\footnotetext{© 2026 IEEE. Personal use of this material is permitted. Permission from IEEE must be obtained for all other uses, in any current or future media, including reprinting/republishing this material for advertising or promotional purposes, creating new collective works, for resale or redistribution to servers or lists, or reuse of any copyrighted component of this work in other works.}

\section{Simultaneous Teachers updates}
\label{sec:sim}
The proposed method requires $N$ teacher models $\mathcal{T}_{1}, \mathcal{T}_{2} ... \mathcal{T}_{N}$ to train independently on $N$ distinct, labelled source datasets $\mathbb{L}_1, \mathbb{L}_2 ... \mathbb{L}_N$. The teacher models parameters are represented by $\Theta_{1},\Theta_{2},...\Theta_{N}$. While the approach is model agnostic, in our case each model is based on standard wav2vec2.0 \cite{baevski2020wav2vec}. Input to the model is the raw waveform, represented by $X$, which get transformed into frame-based features represented by $\mathbf{c_1,...c_{T'}}$, with $T'$ as the time duration of the features. These features form the input to fully connected layers to produce the output. The output of the each teacher is a posterior distribution over tokens, represented by $\hat{\mathbb{P}_i}=[\mathbf{h_1,...h_{T'}}]_i, i \in \{1,...,N\}$. Each model is fine-tuned using CTC \cite{graves2012connectionist} loss represented as follows. 

\begin{equation}
\label{eq:ctc}
    \mathcal{L}_{CTC} = -\sum_{\mathbf{y} \in B^{-1}} \log p(\mathbf{y}|{X})
\end{equation}
where $\mathbf{y}$ represents the labels alignment and $B^{-1}$ contains the set of all possible alignments. 

The student model is represented by $\mathcal{M}$ and has the similar architecture to the teachers. Student model is fine-tuned on data $\mathbb{U}$ whose labels $\hat{\mathbb{L}}$ are generated by teachers. The parameters of the $\mathcal{M}$ are represented by $\Phi$. 

\begin{figure}[t]
  \centering
  \includegraphics[width=1\linewidth]{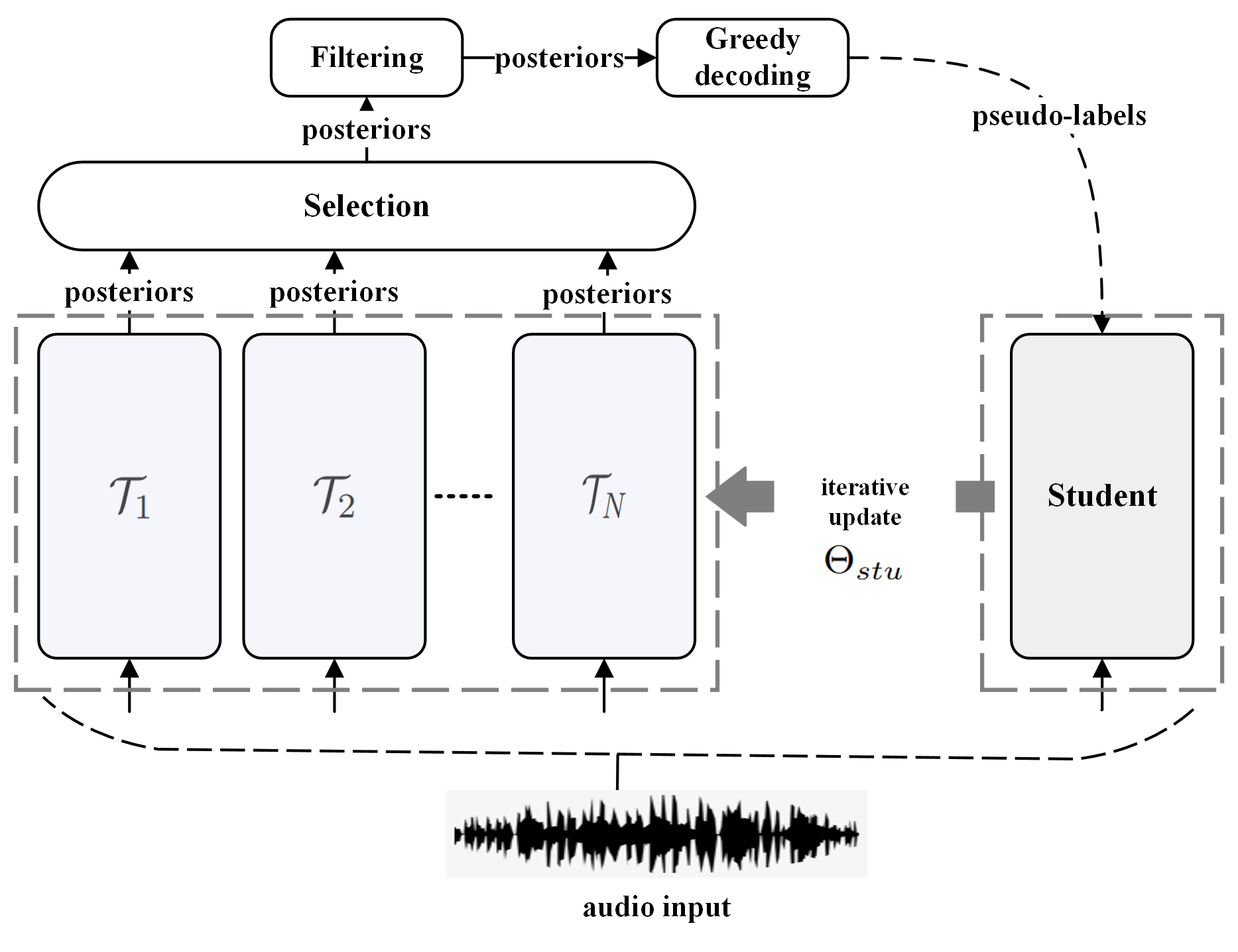}
  \caption{Block diagram for simultaneous teachers update. Given unlabelled audio each teacher model outputs posteriors, one of which is selected and decoded for the student model. Teacher models are simultaneously updated using exponential moving average of the student model weights.}
  \label{fig:model}
\end{figure}

The algorithm starts by training the teacher models on the labelled source data, and by randomly initialising the student model. Unlabelled out-of-domain target data is then passed through the teachers to obtain frame based token posterior distributions. The posteriors acquired from multiple teachers are selected by an elitist selection method \cite{DGKD}, and filtered. During filtering only utterances with high average posterior values are retained, assuming they have lowest error. Greedy decoding is used to convert the posteriors to the pseudo-labels. The student model is update using CTC loss on pseudo-labels acquired from teachers. The teacher models are simultaneously updated after $\Delta$ student updates. Higher the value of $\Delta$, less frequent the teacher model is updated. In contrast to the student model which uses back-propagation to update the weights of the model, each teacher model is updated by an exponential moving average of the student model weights as follows:
\begin{equation}
\label{eq:weightupate}
    \Theta_{i} =\alpha \Phi + (1-\alpha) \Theta_{i}, \ \ i=\{1,2...N\}
\end{equation}
where $\alpha \in [0,1]$ is the proportion of the student model used to update the teacher.
Figure \ref{fig:model} shows the overall system diagram and Algorithm 1 presents the complete algorithm. 

\begin{algorithm}[t]
\label{algo:algo}
\DontPrintSemicolon
  \KwInput{$N$ Labelled source datasets $\mathbb{L}_1, \mathbb{L}_2 ... \mathbb{L}_N$, unlabelled target dataset $\mathbb{U}$, $N$ teacher ASR models $\mathcal{T}_{1}, \mathcal{T}_{2} ... \mathcal{T}_{N}$ and their parameters $\Theta_{1},\Theta_{2},...\Theta_{N}$ student ASR model $\mathcal{M}$ and its parameters $\Phi$, hyper-parameter $\alpha$}
  \KwOutput{$\mathcal{M}$}
  
  \textbf{Initialization:} \\
  \ \ \ Train $\mathcal{T}_{i}$ using $\mathbb{L}_i$, $\forall i \in \{1,2,...N\}$ \\
  \ \ \ Randomly initialise student model $\mathcal{M}$ \\
  \textbf{repeat} \ \ \ \tcp{Loop for epochs}
    \ \ \ \textbf{repeat} \ \ \ \tcp{Loop for batches}
  \ \ \ \ \ \ Draw batch $\bf{B}$ from $\mathbb{U}$ \\
  \ \ \ \ \ \ Input $\bf{B}$ to $N$ teachers to obtain posteriors: $\hat{\mathbb{P}_1}, \hat{\mathbb{P}_2} ... \hat{\mathbb{P}_N}$ \\
  \ \ \ \ \ \ Apply selection and filtering methods to obtain $\hat{\mathbb{P}}$ \\
  \ \ \ \ \ \ Decode $\hat{\mathbb{P}}$ to get pseudo-labels $\hat{\mathbb{L}}$ \\
  \ \ \ \ \ \ Update $\mathcal{M}$ using $\bf{B}$ and $\hat{\mathbb{L}}$ by back-propagation \\
  \ \ \ \ \ \ Update teachers after $\Delta$ iterations: \\
  \ \ \ \ \ \ \ \ \ \ $\Theta_{i} =\alpha \Phi + (1-\alpha) \Theta_{i}, \ \ i=\{1,2...N\}$ \\
  \ \ \ \textbf{until} all batches are done \\
  \textbf{until} maximum epochs \\

\caption{Algorithm for simultaneous teacher updates.}
\end{algorithm}

\subsection{Selection and filtering methods}
The selection method aims to choose the best teacher for each input utterance and filtering decide to either drop that utterance or use its posteriors to train the student. The best teacher is selected for each utterance based on the confidence score \cite{gao2021distilling}, as proposed in \cite{DGKD}. Unlike some offline filtering methods \cite{khurana2021unsupervised, dawalatabad2023unsupervised} that require a model to run inference multiple times making it computationally expensive, our filtering method is online thus requiring single pass. Moreover, both selection and filtering are unsupervised.

The input utterance $X$ is processed by all the teachers to generate output posterior distributions. For each teacher $k$ the maximum posterior value at each time-frame $i$ is computed as:


\begin{equation}
    h_{i,k} = \max \mathbf{h}_{i,k}
\end{equation}
A confidence score of teacher $k$ given any input utterance is calculated by averaging across time, i.e.:


\begin{equation}
\label{eq:avgpost}
    q_k = \frac{1}{T'} \sum_{i=1}^{T'} h_{i,k}
\end{equation}
The teacher model with highest confidence is finally selected as follows:


\begin{equation}
\label{eq4}
    b = \arg \max_k q_k,
\end{equation}
where $b$ represents the identity of the selected teacher.

The subsequent filtering method uses the confidence score of the selected teacher $b$ and applies a threshold to keep or drop that particular utterance. Let $\hat{q}$ be the confidence of the selected teacher, the filtering process keeps the posteriors which meet the threshold $\hat{q} \ge \tau$. Finally, the remaining utterances and the selected posteriors are then used to train the student.

\section{Experiments}

\subsection{Datasets}
Four different datasets are used in the experiments, consisting of WSJ \cite{paul1992design}, LibriSpeech (LS360) \cite{Panayotov2015}, AMI \cite{Carletta2006} and SwitchBoard (SWBD) \cite{Godfrey1992} comprising of read, meeting and conversational telephone speech respectively. In terms of data sizes, AMI consists of 100h, an augmented version of WSJ is 272h, LS360 has 360h and SWBD has 300h. Three datasets (AMI, WSJ and LS360) are used as labelled data to train the teacher models, and SWBD is used as unlabelled target data. The audio of SWBD is upsampled to 16KHz to meet the input requirement of wav2vec2.0 model. To evaluate the performance of the student model the SWBD eval00 set is used, which consists of two subsets from SwitchBoard and CallHome (LDC97S42), represented by SB and CH in the result tables. 


\subsection{Experimental setup}
The experiments make use of three labelled datasets i.e. $N=3$,  one for each of the teacher models and one unlabelled set to update the student model. All models make use of the wav2vec2.0 model pre-trained on LibriVox (LV-60K) \cite{kearns2014librivox}. Wav2vec2.0 models are connected with two fully connected layers and the complete model is fine-tuned using CTC loss. The output of the models consists of 31 tokens including English graphemes and an apostrophe. Each teacher model is fine-tuned independently on each labelled data i.e. AMI, LS360 and WSJ and the student model is fine-tuned on the pseudo-labels generated by the teachers. An out-of-domain 3-gram language model is trained on the AMI, LS360 and WSJ sets. It is used only in evaluation of the student model. 

As discussed in section \ref{sec:sim}, the teacher models are simultaneously updated with the iterative updates of the student model. In this training process three parameters are optimised i.e. the frequency of teachers update $\Delta$, weight of the student parameters $\alpha$ (eq. \ref{eq:weightupate}) and the filtering threshold $\tau$. These parameters were observed to strongly impact the convergence of model training. With the change of one of these, the others need to be adjusted for optimal results. The optimal values for these parameters in our experiments were found to be $\alpha=1e-5, \Delta=40$ and $\tau=0.90$. The relation between these parameters is further discussed in section \ref{sec:results}.

\begin{table*}[t]
    \centering
    \caption{WER(\%) on eval00 test set computed for teachers and student models with and without an out-of-domain (OOD) LM. The shaded cells are results for labelled in-domain SWBD training. The student models consist of a single teacher-student (STS), an ensemble teacher-student (ETS), a multi-stage ensemble teacher-student (METS), KAIZEN and proposed simultaneous teachers update (STU).}
    \label{tab:simteaup}
    \begin{tabular}{l|c|c|c|c|c c c c |c}
        \hline
        \hline
         & \multicolumn{3}{c|}{ \multirow{2}{*}{\bf Teacher models}} & \bf Baseline & \multicolumn{5}{c}{\multirow{2}{*}{\bf Student models}} \\
         & \multicolumn{3}{c|}{} & \bf (labelled data) & \multicolumn{5}{c}{}\\
         \hline
        \multirow{2}{*}{\bf Test sets} & \bf AMI & \bf LS360 & \bf WSJ & \multirow{2}{*}{\bf SWBD} & \multicolumn{5}{c}{\bf SWBD ($\mathcal{M}$)}\\
         & ($\mathcal{T}_{1}$) & ($\mathcal{T}_{2}$) & ($\mathcal{T}_{3}$) & & STS & KAIZEN \cite{manohar2021kaizen} & ETS \cite{DGKD} & METS\cite{ahmad2024progressive} & STU \\
        \hline
        & \multicolumn{9}{c}{\textit{w/o LM}}\\
        \hline
        eval00 & 47.4 & 41.8 & 64.2 & \cellcolor{lightgray} 11.9 & 36.3 & 33.5 & 32.0 & \bf 21.0 & 23.4 \\
        CH & 52.0 & 46.8 & 71.6 & \cellcolor{lightgray} 15.3 & 40.6 & 37.9 & 36.0 & \bf 24.5 & 27.3 \\
        SB & 42.5 & 36.7 & 56.5 & \cellcolor{lightgray} 8.4 & 31.7 & 28.9 & 27.8 & \bf 17.4 & 19.3 \\
        \hline
        & \multicolumn{9}{c}{\textit{w/ LM (OOD)}}\\
        \hline
        eval00 & 44.3 & 38.2 & 61.8 & \cellcolor{lightgray} 10.1 & 31.5 & 29.3 & 26.2 & 19.6 & \bf 18.7 \\
        CH & 49.0 & 43.2 & 69.7 & \cellcolor{lightgray} 12.8 & 35.8 & 33.3 & 30.2 & 23.1 & \bf 22.3 \\
        SB & 39.3 & 33.0 & 53.5 & \cellcolor{lightgray} 7.3 & 27.0 & 25.1 & 22.0 & 16.0 & \bf 15.0\\
        \hline
        \hline
    \end{tabular}
\end{table*}

To compare the proposed method with the state-of-the-art methods, four baselines have been selected. The first is a single teacher-student training (STS) when using the single best teacher model LS360 (measured in performance on the unlabelled data eval00) to train the student model. The second baseline is similar to KAIZEN \cite{manohar2021kaizen} which uses a single teacher model trained on LS360. KAIZEN simultaneously updates the single teacher while training the student model. The third baseline is an ensemble teacher-student (ETS) \cite{DGKD} method. Finally, the fourth baseline is the multi-stage ensemble teacher-student (METS) method \cite{ahmad2024progressive} which trains multiple student models sequentially by considering previously trained model as a teacher. The proposed method is referred to as simultaneous teachers update (STU) and consists of simultaneously updating an ensemble of teacher models along with the training of student model. 

\section{Results and Discussions}
\label{sec:results}
Table \ref{tab:simteaup} show the results of all the experiments. The teacher models are named with respect to the training sets i.e. AMI ($\mathcal{T}_{1}$), LS360 ($\mathcal{T}_{2}$) and WSJ ($\mathcal{T}_{3}$). All models are evaluated on eval00 and its two subsets CallHome (CH) and SwitchBoard (SB). The table shows that among three teacher models the best performing teacher model is LS360 with an eval00 WER of 41.8\% (w/o LM) and 38.2\% (w/ LM). The second and third best teacher models are AMI and WSJ. In single teacher experiments (STS \& KAIZEN), the LS360 teacher model is the basis. WER for the baseline SWBD model, trained on labelled data is presented in shaded cells. This baseline show minimum WER any model can achieve on SWBD eval00 test set. 

\begin{figure}[t]
     \centering
     \begin{subfigure}[b]{0.49\linewidth}
         \centering
         \includegraphics[width=1.0\linewidth]{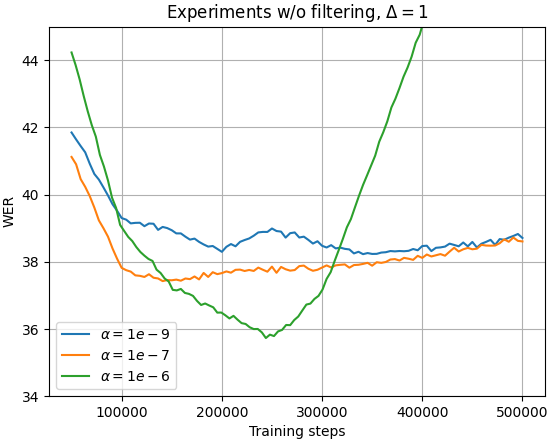}
         \caption{}
         \label{a}
     \end{subfigure}
     \hfill
     \begin{subfigure}[b]{0.49\linewidth}
         \centering
         \includegraphics[width=1.0\linewidth]{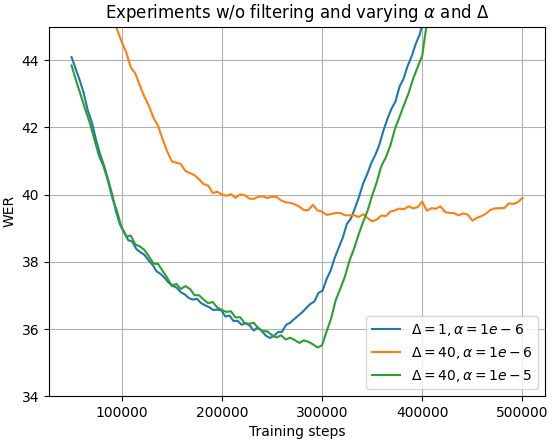}
         \caption{}
         \label{b}
     \end{subfigure}
     \hfill
     \begin{subfigure}[b]{0.49\textwidth}
         \centering
         \includegraphics[width=0.5\textwidth]{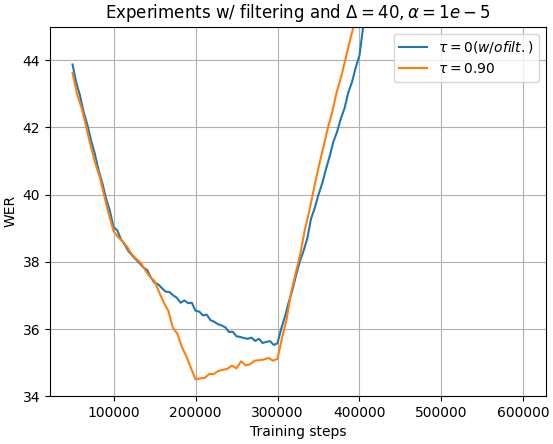}
         \caption{}
         \label{c}
     \end{subfigure}
        \caption{WER of validation set during training of the student model. The figures show WERs under varying parameters $\alpha, \Delta$ and $\tau$. 
        }
        \label{fig:wers}
\end{figure}

The STS method shows the advantage of using the T/S training method with a single best teacher, which yields better results than any teacher on the OOD eval00 test set, with a WER of 36.3\%. Compared to the LS360 model, the STS method is having an improvement of 5.5\% absolute. On the complete eval00 set the KAIZEN method is 2.8\% absolute better than STS, with a WER of 33.5\% (w/o LM) and 29.3\% (w/ LM). Similar improvements are found for the CH and SB subsets. Updating the teacher model helps the student model to get improved labels in the next iteration, thereby achieving better results in the end. Results produced with the ensemble teacher-student (ETS) method show the advantage of using an ensemble of teachers. Even though, WSJ and AMI models perform worse than the LS360 one,  the selection method leads to an overall lower error and therefore achieves better results. Compared to KAIZEN with a WER of 33.5\% (w/o LM) and 29.3\%(w/ LM) on eval00, ETS yields 32.0\% (w/o LM) and 26.2\% (w/ LM) WER, with the highest absolute difference of about 3.1\%. The ETS results were further improved by multi-stage training (METS) \cite{ahmad2024progressive} , i.e. by training sequence of students iteratively, using the previous student as a teacher. METS significantly improved the WERs from 32.0\% to 21.0\% (w/o LM), and 26.2\% to 19.6\% (w/ LM). The METS method is computationally expensive because each student model goes through full training iterations until it becomes a teacher to produce labels for the next stage. In METS training, an LM was included, and beam search was used to produce labels at each stage. The proposed STU method reduces this complexity by using a single student and greedy decoding, but additionally updates the teachers. The teacher model updates in STU is a simple moving average weight update method. 

Compared to STS, the proposed STU method has up to 12.9\% better absolute WER for eval00, with and without LM. The best improvement is shown on the CH subset, improving the WER from 35.5\% to 22.3\% (w/ LM). Compared to the KAIZEN method, the improvement is up to 10.6\% (w/o LM) and 11\% (w/ LM) on the CH subset. On eval00, the improvement for STU is 10.1\% WER absolute (w/o LM) and 10.6\% (w/ LM). Compared to the ensemble teacher method ETS, STU gains are 8.6\% (w/o LM) and 7.5\% (w/ LM) respectively, with  similar differences for CH and SB subsets. In comparison to METS, the STU method improves by 1\% WER, when the decoding is performed using an LM. It should be noted that in contrast to  METS, the STU method apply greedy decoding at the output of teachers to generate pseudo-labels for the student model training. The improvement for STU can be further improved by using an LM during training to get better pseudo-labels for the student. However, this increases the training time significantly, therefore not included in the experiment. 

Figure \ref{fig:wers} shows the WER convergence for the STU model with different values of $\alpha, \Delta$ and $\tau$ during training. In Figure \ref{a}, different values of $\alpha$ have been used to observe its effect during training, while keeping $\Delta=1$. It can be observed that for $\alpha=10^{-9}$, the student models do not tend to diverge because the teacher model is only updated with a small fraction from the student model. However, the WER stays higher compared to other graphs. For $\alpha=10^{-7}$, the model shows lower WER but tends to diverge gradually. Further increasing the value of $\alpha$ to $10^{-6}$ achieves the lowest WER but training becomes unstable after 250k training steps. This is the lowest WER, and a further increase of $\alpha$ does not lead to convergence. Figure \ref{b} shows that when $\Delta$ is increased to $40$, with $\alpha=10^{-6}$ the training becomes stable but the WER stays higher. The value for $\alpha$ is further increased for lower WER. The lowest error is finally obtained with $\Delta=40 $ and $\alpha=10^{-5}$. Similarly, after selecting $\alpha$ and $\Delta$ the value of $\tau$ is chosen. The parameter $\tau$ is used to filter utterances with potentially high error. If fewer erroneous utterances are chosen then the model should also have better performance. This is shown in Figure \ref{c} . 
With a suitable value of $\tau$ the model yields the lowest error rate while $\alpha$ and $\Delta$ are fixed to $40$ and $1e-5$.

\section{Conclusion}

This paper proposed a novel simultaneous teachers update method for ensemble T/S training to improve unsupervised domain adaptation. Experiments first show the advantage of using an ensemble of teachers in unsupervised domain adaptation, and further gain when simultaneously updating teachers. The teacher model updates are shown to be an inexpensive method, utilising an exponential moving average of the student model weights. The proposed method outperforms all baselines including ensemble, multi-stage and iterative methods. For future work, we aim to propose mechanism to control possible model collapse specifically for out-of-domain data since the previously suggested controlling techniques for in-domain data were found not to work in our scenario. 

\section{Acknowledgements}
This work was partially supported by the LivePerson center in the Speech and Hearing Group at the University of Sheffield, UK.

\bibliographystyle{IEEEtran}
\bibliography{refs}

\end{document}